\documentclass[]{interact}

\usepackage{amsmath,amssymb,amscd,latexsym,dsfont,wasysym,bbm}
\usepackage{float}
\usepackage{color}
\usepackage{graphicx,subfigure}
\usepackage{multirow,multicol} 
\usepackage{verbatim}
\usepackage{psfrag}
\usepackage{comment}
\usepackage{makeidx}
\usepackage[]{algorithm}
\usepackage{algorithmicx}
\usepackage{algpseudocode}
\usepackage{cases}
 \usepackage{setspace}
\usepackage{pgfplots} 
\newcommand{\tr}[1]{\textrm{#1}}
\newcommand{\mr}[1]{\mathrm{#1}}
\newcommand{\tnr}[1]{{\textnormal{#1}}}

\newcommand{\mc}[1]{\mathcal{#1}}
\newcommand{\mf}[1]{\mathsf{#1}}
 
\newcommand{\ms}[1]{\mathds{#1}}

\newcommand{\ov}[1]{\overline{#1}}

\newcommand{\bI}{\boldsymbol{I}}

\newcommand{\bx}{\boldsymbol{x}}

\newcommand{\bone}{\boldsymbol{1}}

\newcommand{\btheta}{\boldsymbol{\theta}}

\newcommand{\bxi}{\boldsymbol{\xi}}

\newcommand{\secref}[1]{Sec.~\ref{#1}}
\newcommand{\appref}[1]{Appendix~\ref{#1}}

\newcommand{\tabref}[1]{Table~\ref{#1}}

\newcommand{\ie}{i.e.,~} 		
\newcommand{\eg}{e.g.,~}	

\newcommand{\argmax}{\mathop{\mr{argmax}}}
\newcommand{\argmin}{\mathop{\mr{argmin}}}

\newcommand{\set}[1]{\{#1\}}
\newcommand{\SET}[1]{\left\{#1\right\}}

\newcommand{\cd}{\cdot}
\newcommand{\ld}{\ldots}

\newcommand{\e}{\mr{e}}

\newcommand{\PR}[1]{\Pr\SET{#1}}       	

\newcommand{\IND}[1]{\ms{I}\big[{#1}\big]}   	
\newcommand{\Ex}{\ms{E}}     			
\newcommand{\Var}{\ms{V}\tr{ar}}     			
\newcommand{\T}{^{\top}}            		
\newcommand{\dd}{\,\mr{d}}             		

\newcommand{\mcT}{\mc{T}}

\newcommand{\mcY}{\mc{Y}}

\newcommand{\mfm}{\mf{m}}

\newcommand{\mfA}{\mf{A}}
\newcommand{\mfD}{\mf{D}}

\newcommand{\mfH}{\mf{H}}

\newcommand{\matH}{\tnr{\textbf{H}}}

\usepackage{colortbl}
\newcommand{\ccol}{\cellcolor{blue!25}} 

\pgfplotsset{compat=1.12}
\usetikzlibrary{positioning,shadows,shapes,fit,arrows,backgrounds}

\tikzstyle{rect_my} = [draw, rectangle, minimum width=2cm, text width=1.8cm, fill=gray!15, 
  text centered,  minimum height=.9cm]
\tikzstyle{square_my} = [draw, rectangle, minimum width=1cm, text width=0.8cm, fill=gray!15, 
  text centered,  minimum height=.9cm]
\tikzstyle{square_my_graph} = [draw, rectangle, minimum width=1.2cm, text width=1cm, fill=gray!15, 
  text centered,  minimum height=1.2cm]
\tikzstyle{circle_my} = [draw, circle, minimum width=1cm, text width=0.8cm, fill=gray!15, 
  text centered,  minimum height=.9cm]
\tikzstyle{circle_my_graph} = [draw, circle, minimum width=1.1cm, text width=.8cm, fill=gray!15, 
  text centered]
\tikzstyle{cloud_my} = [draw, shape=cloud, minimum width=1cm, text width=0.8cm, fill=gray!15, 
  text centered,  minimum height=1cm]

\tikzstyle{point_my} = [draw=none, minimum width=0cm, text width=0cm, fill=none, 
  text centered,  minimum height=0cm]    
\tikzstyle{line_my} = [draw, -latex]    
\tikzstyle{box_my}=[draw, minimum size=2em, text width=4.5em, text centered]
\tikzstyle{bigbox_my}=[draw, inner sep=15pt]
\tikzstyle{arrow_my} = [thick,->,>=stealth]
\tikzstyle{noarrow_my} = [thick,-,=>stealth]

\newcommand{\data}[1]{{\color{black} #1}}

\usepackage[acronym,nonumberlist]{glossaries} 

\usepackage[natbibapa,nodoi]{apacite}
\usepackage{hyperref}
\usepackage{empheq}
\usepackage{verbatim}
\usepackage{graphicx}

\newacronym[\glsshortpluralkey=PDFs,\glslongpluralkey=probability density functions]{pdf}{PDF}{probability density function}
\newacronym[\glsshortpluralkey=CDFs,\glslongpluralkey=cumulative density functions]{cdf}{CDF}{cumulative density function}
\newacronym[\glsshortpluralkey=CCDFs,\glslongpluralkey=complementary cumulative density functions]{ccdf}{CDF}{complementary cumulative density function}
\newacronym[\glsshortpluralkey=PMFs,\glslongpluralkey=probability mass functions]{pmf}{PMF}{probability mass function}
\newacronym[]{lhs}{l.h.s.}{left-hand side}
\newacronym[]{rhs}{r.h.s.}{right-hand side} 

\newacronym[]{bicm}{BICM}{bit-interleaved coded modulation}
\newacronym[]{bicmid}{BICM-ID}{BICM with iterative demapping}
\newacronym[]{cm}{CM}{coded modulation}
\newacronym[]{tcm}{TCM}{trellis-coded modulation}
\newacronym[]{mlc}{MLC}{multi-level coding}
\newacronym[]{pam}{PAM}{pulse amplitude modulation}
\newacronym[]{bpsk}{BPSK}{binary phase shift keying}
\newacronym[]{qam}{QAM}{quadrature amplitude modulation}
\newacronym[]{16qam}{16-QAM}{16-points quadrature amplitude modulation}
\newacronym[]{psk}{PSK}{phase shift keying}
\newacronym[\glsshortpluralkey=LLRs,\glslongpluralkey=logarithmic likelihood ratios]{llr}{LLR}{logarithmic likelihood ratio}
\newacronym[]{oc}{OC}{operating characteristic}
\newacronym[]{dmp}{DMP}{discretized message passing}
\newacronym[]{mp}{MP}{message passing}
\newacronym[]{ep}{EP}{expectation propagation}
\newacronym[\glsshortpluralkey=MIs,\glslongpluralkey=mutual informations]{mi}{MI}{mutual information}
\newacronym[\glsshortpluralkey=GMIs,\glslongpluralkey=generalized mutual informations]{gmi}{GMI}{generalized mutual information}
\newacronym[]{eesm}{EESM}{exponential effective-SNR-mapping}
\newacronym[]{bicm-gmi}{BICM-GMI}{BICM generalized mutual information}
\newacronym[]{awgn}{AWGN}{additive white Gaussian noise}
\newacronym[]{bsc}{BSC}{binary symetric channel}
\newacronym[]{amc}{AMC}{adaptive modulation and coding}
\newacronym[]{csi}{CSI}{channel state information}
\newacronym[]{cqi}{CQI}{channel quality indicator}
\newacronym[]{kl}{KL}{Kullback-Leibler}
\newacronym[]{cmm}{CMM}{circular moment matching}
\newacronym[]{ga}{GA}{Gaussian approximation}

\newacronym[]{sp}{SP}{set-partitioning}
\newacronym[]{gsm}{GSM}{global system for mobile communications}
\newacronym[]{edge}{EDGE}{enhanced data rates for GSM evolution}
\newacronym[]{3gpp}{3GPP}{3rd generation partnership project}
\newacronym[]{umts}{UMTS}{Universal Mobile Telecommunication System}
\newacronym[]{lte}{LTE}{Long Term Evolution}
\newacronym[]{dvb}{DVB}{digital video broadcasting}
\newacronym[]{fdd}{FDD}{Frequency Division Duplexing}

\newacronym[\glsshortpluralkey=CCs,\glslongpluralkey=convolutional codes]{cc}{CC}{convolutional code}
\newacronym[\glsshortpluralkey=PCCCs,\glslongpluralkey=parallel concatenated convolutional codes]{pccc}{PCCC}{parallel concatenated convolutional code}
\newacronym[\glsshortpluralkey=TCs,\glslongpluralkey=turbo codes]{tc}{TC}{turbo code}
\newacronym{ldpc}{LDPC}{low-density parity-check}
\newacronym[]{ofdm}{OFDM}{orthogonal frequency-division multiplexing}
\newacronym[]{bep}{BEP}{bit-error probability}
\newacronym[]{wep}{WEP}{word-error probability}
\newacronym[]{sep}{SEP}{symbol-error probability}
\newacronym[]{pep}{PEP}{pairwise-error probability}
\newacronym[]{ttcm}{TTCM}{turbo-trellis coded modulation}
\newacronym[]{uep}{UEP}{unequal error protection}
\newacronym[\glsshortpluralkey=CENCs,\glslongpluralkey=convolutional encoders]{cenc}{CENC}{convolutional encoder}
\newacronym[]{mimo}{MIMO}{multiple-input multiple-output}
\newacronym[\glsshortpluralkey=SNRs,\glslongpluralkey=signal-to-noise ratios]{snr}{SNR}{signal-to-noise ratio}
\newacronym[\glsshortpluralkey=SINRs,\glslongpluralkey=the signal-to-interference-plus-noise ratios]{sinr}{SINR}{the signal-to-interference-plus-noise ratio}
\newacronym[]{msb}{MSB}{most-significative bit}
\newacronym[]{bcjr}{BCJR}{Bahl--Cocke--Jelinek--Raviv}
\newacronym[]{cbc}{CBC}{Colavolpe--Barbieri--Caire}
\newacronym[]{skr}{SKR}{Shayovitz--Kreimer--Raphaeli}
\newacronym[\glsshortpluralkey=SEDs,\glslongpluralkey=squared Euclidean distances]{sed}{SED}{squared Euclidean distance}
\newacronym[\glsshortpluralkey=EDs,\glslongpluralkey=Euclidean distances]{ed}{ED}{Euclidean distance}
\newacronym[\glsshortpluralkey=MEDs,\glslongpluralkey=minimum Euclidean distances]{med}{MED}{minimum Euclidean distance}
\newacronym[]{core}{CoRe}{constellation rearrangement}
\newacronym[]{msd}{MSD}{multistage decoding}
\newacronym[]{pdl}{PDL}{parallel decoding of the individual levels}
\newacronym[\glsshortpluralkey=GCs,\glslongpluralkey=Gray codes]{gc}{GC}{Gray code}
\newacronym[]{brgc}{BRGC}{binary-reflected Gray code}
\newacronym[]{nbc}{NBC}{natural binary code}
\newacronym[]{fbc}{FBC}{folded-binary code}
\newacronym[]{bsgc}{BSGC}{binary semi-Gray code}
\newacronym[]{msp}{MSP}{modified set-partitioning}
\newacronym[]{ssp}{SSP}{semi set-partitioning}
\newacronym[]{fhd}{FHD}{free Hamming distance}
\newacronym[]{mfhd}{MFHD}{maximum free Hamming distance}
\newacronym[]{ods}{ODS}{optimal distance spectrum}
\newacronym[]{iud}{i.u.d.}{independent and uniformly distributed}
\newacronym[]{ud}{u.d.}{uniformly distributed}
\newacronym[]{iid}{i.i.d.}{independent, identically distributed}
\newacronym[]{ami}{AMI}{accumulated mutual information}
\newacronym[]{bico}{BICO}{binary-input continuous-output}
\newacronym[]{gh}{GH}{Gauss--Hermite}
\newacronym[]{gum}{GUM}{Gaussian--uniform mixture}

\newacronym[\glsshortpluralkey=BSs,\glslongpluralkey=base-stations]{bs}{BS}{base-station}
\newacronym[\glsshortpluralkey=MSs,\glslongpluralkey=mobile-stations]{ms}{MS}{mobile-stations}

\newacronym[]{phy}{PHY}{physical layer} 
\newacronym[]{rlc}{RLC}{Radio-Link control} 
\newacronym[]{ran}{RAN}{Radio Access Network} 
\newacronym[]{llc}{LLC}{logical link control} 
\newacronym[]{tcp}{TCP}{transmission control protocol} 
\newacronym[]{mac}{MAC}{media access control} 
\newacronym[]{fft}{FFT}{fast Fourier transform} 
\newacronym[]{ft}{FT}{Fourrier transform}
\newacronym[]{cf}{CF}{characteristic function} 
\newacronym[]{mgf}{MGF}{moment generating function} 
\newacronym[]{ee}{EE}{energy efficiency} 
\newacronym[]{eb}{EB}{energy per bit}
\newacronym[]{kkt}{KKT}{Karush--Kuhn--Tucker} 
\newacronym[]{mcs}{MCS}{modulation/coding scheme} 
\newacronym[]{fec}{FEC}{forward error correction}
\newacronym[]{arq}{ARQ}{automatic repeat request}
\newacronym[]{harq}{HARQ}{hybrid ARQ}
\newacronym[]{tarq}{TARQ}{truncated HARQ}
\newacronym[]{ir}{IR}{incremental redundancy}
\newacronym[]{rpr}{RR}{repetition redundancy}
\newacronym[]{rrharq}{RR-HARQ}{repetition redundancy HARQ}
\newacronym[]{irharq}{IR-HARQ}{incremental redundancy HARQ}
\newacronym[]{ack}{ACK}{positive acknowledgment}
\newacronym[]{nack}{NACK}{negative acknowledgment}
\newacronym[]{hol}{HoL}{head of the line}
\newacronym[]{crc}{CRC}{cyclic redundancy check}
\newacronym[]{dp}{DP}{dynamic programming}
\newacronym[]{gp}{GP}{geometric programming}
\newacronym[]{per}{PER}{packet error rate}
\newacronym[]{ber}{BER}{bit error rate}
\newacronym[]{op}{OP}{outage probability}
\newacronym[]{spa}{SPA}{saddle-point approximation}
\newacronym[]{mrc}{MRC}{maximum ratio combining}
\newacronym[]{mdp}{MDP}{Markov decision process}
\newacronym[]{lp}{LP}{linear programming}
\newacronym[]{pomdp}{POMDP}{partially observable Markov decision process}
\newacronym[]{psimdp}{PSI-MDP}{partial state information Markov decision process}
\newacronym[]{scpp}{SCPP}{stochastic shortest path problem}

\newacronym[]{forw}{frwd}{forward}
\newacronym[]{feed}{fdbk}{feedback}

\newacronym[]{mm}{MM-HARQ}{multi-message HARQ}
\newacronym[]{xp}{XP-HARQ}{cross-packet HARQ}
\newacronym[]{ts}{TS}{time-sharing}
\newacronym[]{sc}{SC}{superposition coding}
\newacronym[]{sbrq}{SBRQ}{systematic backward retransmission}
\newacronym[]{brq}{BRQ}{backward retransmission}
\newacronym[]{lharq}{L-HARQ}{layer-coded HARQ}
\newacronym[]{anlharq}{AoN-HARQ}{all-or-none L-HARQ}
\newacronym[]{vlharq}{VL-HARQ}{variable-length HARQ}

\newacronym[]{pp}{PPP}{point process}
\newacronym[]{ppp}{PPP}{Poisson point process}

\newacronym[]{fide}{FIDE}{F\'ed\'eration Internationale des \'Echecs}
\newacronym[]{fifa}{FIFA}{F\'ed\'eration Internationale de Football Association}
\newacronym[]{fivb}{FIVB}{F\'ed\'eration Internationale de Volleyball}
\newacronym[]{epl}{EPL}{English Premier League}
\newacronym[]{nhl}{NHL}{National Hockey League}
\newacronym[]{nfl}{NFL}{National Football League}

\newacronym[]{sg}{SG}{stochastic gradient}
\newacronym[]{lms}{LMS}{least mean squares}
\newacronym[]{rls}{RLS}{recursive least squares}
\newacronym[]{vss}{VSS}{variable step-size}
\newacronym[]{hfa}{HFA}{home-field advantage}
\newacronym[]{ha}{HA}{home advantage}
\newacronym[]{mov}{MOV}{margin of victory}
\newacronym[]{ac}{AC}{Adjacent Categories}
\newacronym[]{cl}{CL}{Cumulative Link}
\newacronym[]{rps}{RPS}{Ranked Probability Score}
\newacronym[]{mse}{MSE}{Mean Square Error}
\newacronym[]{mmse}{MMSE}{Minimum Mean Square Error}
\newacronym[]{rmse}{RMSE}{Root Mean Squares Error}
\newacronym[]{map}{MAP}{maximum a posteriori}
\newacronym[]{ml}{ML}{maximum likelihood}
\newacronym[]{loo}{LOO}{leave-one-out}
\newacronym[]{alo}{ALO}{approximate leave-one-out}

\newacronym[]{svd}{SVD}{singular value decomposition}

\newacronym[]{skf}{SKF}{Simplified Kalman Filter}
\newacronym[]{vskf}{vSKF}{\emph{vector-covariance} Simplified Kalman Filter}
\newacronym[]{sskf}{sSKF}{\emph{scalar-covariance} Simplified Kalman Filter}
\newacronym[]{fskf}{fSKF}{\emph{fixed-variance} Simplified Kalman Filter}
\newacronym[]{kf}{KF}{Kalman Filter}
\newacronym[]{gelo}{G-Elo}{Generalized Elo}

\newacronym[]{tpb}{TPB}{tensor-product-basis}

\begin{document}

\title{FIFA ranking: \\ Evaluation and path forward}
\author{Leszek Szczecinski and Iris-Ioana Roatis
\thanks{L.~Szczecinski  is with Institut National de la Recherche Scientifique, Montreal, Canada [e-mail: Leszek.Szczecinski@inrs.ca].}
\thanks{I.-I.~Roatis is with Imperial College, London, UK [e-mail: iris-ioana.roatis18@imperial.ac.uk].}
}

\maketitle

\begin{abstract}
    In this work we study the ranking algorithm used by \acrfull{fifa}; we analyze the parameters it currently uses, show the formal probabilistic model from which it can be derived, and optimize the latter. In particular, analyzing the games since the introduction of the algorithm in 2018, we conclude that the game's ``importance'' (as defined by \acrshort{fifa}) used in the algorithm is counterproductive from the point of view of the predictive capability of the algorithm. We also postulate the algorithm to be rooted in the formal modelling principle, where the Davidson model proposed in 1970 seems to be an excellent candidate, preserving the form of the algorithm currently used. The results indicate that the predictive capability of the algorithm is notably improved by using the \acrfull{hfa} and the explicit model for the draws in the game. Moderate, but notable improvement may be attained by introducing the weighting of the results with the goal differential, which although not rooted in a formal modelling principle, is compatible with the current algorithm and can be tuned to the characteristics of the football competition.
\end{abstract}

\section{Introduction}
 In this work paper we evaluate the algorithm used by \gls{fifa} to rank the international men teams. We also propose and study simple modifications to improve the prediction capability of the algorithm.

We are motivated by the fact that the rating and ranking are important elements of sport competitions and the surrounding entertainment environments. The rating consists in assigning the team/player a number, often referred to as ``skills'' or ``strengths''; the ranking is obtained by sorting these numbers and is also referred to as ``power ranking".

The rating has an informative function providing fans and profane observers with a quick insight into the relative strength of the teams. For example, the press is often interested in the ``best'' teams or the national team reaching some records position in the ranking.

More importantly, the ranking leads to consequential decisions such as a) the seeding, \ie defining which teams play against each other in the competitions (\eg used to establish the composition of the groups in the qualification rounds of the \gls{fifa} World Cup),  b) the promotion/relegation (\eg determining which teams  move between the \gls{epl} and the English Football League Championship, or teams which move between the Nations Leagues groups), or c) defining the participants in the prestigious (and lucrative) end-of-the-season competitions (such as Champions League in European football, Stanley Cup series in \gls{nhl}). 

Most of the currently used sport ratings are based on counting of wins/loses (and draws, when applicable) but in some cases the sport-governing bodies moved beyond these simple methods and implemented more sophisticated rating algorithms where the rating levels attributed to the teams are meant to represent the skills.

In particular, \gls{fifa} started a new ranking/rating algorithm in 2018, where the rating level (skills) assigned to the teams are calculated from the game outcome, of course, but also from the skills of the teams before the game. The resulting rating algorithm has a virtue of being simple and defined in a (mostly) transparent manner.

The main objective of this work is to analyze the \gls{fifa} ranking using the statistical modelling methodology. Considering that the football association is, by any measure, the most popular sport in the world, it has a value in itself and follows the line of works which analyzed the past strategies of ranking used by  \gls{fifa}, \eg \citep{Lasek13}, \citep{Ley19}. However, the approach we propose can be applied to evaluate other algorithms as well, such as the one used by \gls{fivb}, \citep{fivb_rating}.

In this work we will:
\begin{itemize}
    \item 
Derive the \gls{fifa} algorithm from the first principles. In particular, we will define the probabilistic model underlying the algorithm and identify the estimation method used to estimate the skills. 
    \item
Assess the relevance of the parameters used in the current algorithms. In particular we will evaluate the role played by the change of the adaptation step according to the game's importance (as defined by \gls{fifa}).
    \item
Optimize the parameters of the proposed model. As a result, we derive an algorithm which is equally as simple as the \gls{fifa}'s one, but allows us to improve the prediction of the games' results.
    \item
Propose the modifications of the algorithm which take into account the goals differential, also known as the \gls{mov}. We consider legacy-compliant algorithms and a new version of the rating.
\end{itemize}

Our work is organized as follows. In \secref{Sec:FIFA.rating} we describe the \gls{fifa} algorithm in the framework which simplifies the manipulation of the models and the evaluation of the results. This is also where we clarify the data origin and make a preliminary evaluation of the relevance of the game's importance parameters currently used to control the size of the adaptation step. The algorithm is then formally derived in \secref{Sec:derivation.and.batch} where we also discuss the evaluation of the results and the batch estimation approach we use. The incorporation of the \gls{mov} into the rating is evaluated in \secref{Sec:MOV.general} using two different strategies. In \secref{Sec:on-line} we return to the on-line rating, evaluating and re-optimizing the proposed algorithms, pointing out to the role of the scale, and commenting on the elements of the \gls{fifa} algorithm (the shootout/knockout rules) which seem to be introduced in an ad-hoc manner for which the models are not specified. We conclude the work in \secref{Sec:Conclusions} summarizing our findings and in \secref{Sec:Recommedations} we make en explicit list of recommendations which may be introduced to improve on the current version of the \gls{fifa} algorithm.

\section{FIFA ranking algorithm}\label{Sec:FIFA.rating}

We consider the scenario where there is a total of $M$ teams playing agains each other in the games indexed with $r=1,\ld,T$, where $T$ is the number of games in the observed period. \gls{fifa} ranks $M=210$ international teams and, between June 4, 2018 and \data{October 16, 2021, there were $T=2964$} games \gls{fifa}-recognized games.

Let us denote the skill of the team $m=1, \ld, M$ before the game $t$ as $\theta_{t,m},~ t\in\mcT=\set{1,\ld, T}$ which are gathered in a vector $\btheta_t=[\theta_{t,1},\ld,\theta_{t,M}]\T$, where $(\cd)\T$ denotes the transpose. The home and the away teams are denoted by $i_t$ and $j_t$ respectively. 

The game results $y_t\in\mcY$ are ordinal variables, where the elements of $\mcY=\set{\mfH,\mfD, \mfA}$  represent the win of the home team ($y_t=\mfH$), the draw ($y_t=\mfD$), and the win of the away team ($y_t=\mfA$). These ordinal variables are often transformed into the numerical \emph{scores} $\check{y}_t=\check{y}(y_t)$: $\check{y}(\mfA)=0$, $\check{y}(\mfD)=0.5$ and $\check{y}(\mfH)=1$.

The basic rules of \gls{fifa}'s rating for a team $m$ which plays in the $t$-th game are defined as follows
\begin{align}\label{FIFA.basic.rules}
    \theta_{t+1,m} & \leftarrow \theta_{t,m} + I_{c_t} \delta_{t,m}\\
    \label{FIFA.basic.delta}
    \delta_{t,m} & = \check{y}_{t,m} - F\big( \textstyle \frac{z_{t,m}}{s}\big) \\
    \label{FIFA.logistic}
    F(z) & = \frac{1}{1+10^{-z}}\\
    z_{t,m} &= \theta_{t,m} - \theta_{t,n}
\end{align}
where $s=600$ is the scale\footnote{The role of the scale is to ensure that the values of the skills $\theta_{t,m}$ are situated in a visually comfortable range; it can be also used when changing the rating algorithm, as is discussed in \secref{Sec:Scale.adjustment}.}, $n$ is the index of the team opposing the team $m$ in the $t$-th game, $\check{y}_{t,m}$ is the ``subjective''  score of the team $m$ (for the home team, $m=i_t$, $\check{y}_{t,m}=\check{y}_t$, and for the away team, $m=j_t$, $\check{y}_{t,m}=1-\check{y}_t$). The result produced by the logistic function, $F(z_{t,m}/s)$ in \eqref{FIFA.basic.delta} is referred to as the \emph{expected score}. 

When the team $m$ does not play, its skills do note change, \ie $\theta_{t+1,m}\leftarrow\theta_{t,m}$.

Since $|\delta_{t,m}|\le 1$, $I_{c_t}$ is the maximum allowed update step, where $c_t$ is the game category (or game ``importance'') and we can decompose $I_c$ into two components
\begin{align}\label{define.I_c}
    I_c = K \xi_c,
\end{align}
where $K=5$ and $\xi_c$ is a category-dependent adjustment as defined in \tabref{Tab:importance_levels}.

\begin{table}[th]
    \centering
    \begin{tabular}{c|c|c|c|c}
$c$ & $I_c$ & $\xi_c$ & Description & Number\\
\hline
0 & 5 & 1  & Friendlies outside International Match Calendar windows  & 436\\
1 & 10 & 2  & Friendlies during International Match Calendar windows  & 583\\
2 & 15 & 3  & Group phase of Nations League competitions  & 347\\
3 & 25 & 5  & Play-offs and finals of Nations League competitions & 84\\ 
4 & 25 & 5  & Qualifications for Confederations/World Cup finals & 1189\\
5 & 35 & 7  & Confederation finals up until the QF stage  & 209\\ 
6 & 40 & 8  & Confederation finals from the QF stage onwards & 52\\  
7 & 50 & 10 & World Cup finals up until QF stage    & 56\\
8 & 60 & 12 & World Cup finals from QF stage onwards & 8
\end{tabular}
    \caption{Categories, $c$ of the game and the corresponding update steps $I_c=K \xi_c$, \citep{fifa_rating}, where $K=5$ and $\xi_c=I_c/I_0$. The number of the observed categories between June 4, 2018 and \data{October 16, 2021 is also given (total number of games is $T=2964$)}.}
    \label{Tab:importance_levels}
\end{table}

The basic equation governing the change of the skills in \eqref{FIFA.basic.delta} is next supplemented with the following rules:
\begin{itemize}
    \item \emph{Knockout rule}: in the knockout stage of any competition (which follows the group stage), instead of \eqref{FIFA.basic.delta} we use 
    \begin{align}\label{knockout.rule}
        \delta_{t,m} & \leftarrow \max\set{0, \delta_{t,m}}
    \end{align}
    which guarantees that no points are lost by teams moving out of the group stage.
    \item \emph{Shootout rule}: 
    If the team $m$ wins the game in the shootouts we use 
    \begin{align}\label{shootout.rule}
        \check{y}_{t,m}&\leftarrow 0.75,\quad
        \check{y}_{t,n}\leftarrow 0.5,
    \end{align}
    where $n$ is the index of the team which lost.
    
    This rule, however, does not apply in two-legged qualification games if the shootout is required to break the tie.
\end{itemize}

We will discuss the effect of the knockout rule later and here we only point out to the fact that by applying the shootout/knockout we always increase $\delta_{t,m}$. Thus, while the basic rules \eqref{FIFA.basic.rules}-\eqref{FIFA.basic.delta} guarantee that the teams ``exchange'' the points so the total number of points stays constant, \ie $\sum_{m=1}^M\theta_{t,m}=\sum_{m=1}^M\theta_{t+1,m}$ (this is a well-known property of the Elo rating algorithm, \citep{Elo78_Book}), the shootout/knockout rules increase the total number of points which causes an ``inflation" of the rating. In fact, in the considered period, there were \data{124 games where the shootout rule, the knockout rule, or both were applied and this increased the total score by $1739$ points} (with the initial total being $254680$).

The rating we described is published by \gls{fifa} since August 2018, roughly on a per-month basis. The algorithm was initialized on the June 4, 2018, with the initialization values $\btheta_0$ based on the previous rating system.

To run the algorithm, we need to know the initialization $\btheta_0$, the presence of conditions which trigger the use of the knockout/shootout rules, and most importantly, the category/importance of each game, $c_t$. These elements are not officially published so we use here the unofficial data shown in \citet{football_rankings} which keeps track of the \gls{fifa} rating since June 2018. Using it,  we were able to reproduce the ratings $\btheta_t$ with a precision of fractions of rating points which gives us confidence that the categories of the games are assigned according to the \gls{fifa} rules.\footnote{Information provided by \citet{football_rankings} is highly valuable because it is far from straightforward to verify which games are included in the rating and what their importance $I_c$ is. In particular, the games in the same tournament can be included or excluded from the rating and in some cases the changes may be done retroactively complicating further the understanding of the rating results. For example, we had to deal with two minor exceptions:
\begin{itemize}
    \item 
We recognized the victory of Guyana (GUY) over Barbados (BRB) in the game played on Sept. 6, 2019 already on the date of the game, while in the \gls{fifa} rating, the draw was originally registered and the GUY's victory was recognized only later, when BRB was disqualified for having fielded an ineligible player.
    \item 
We removed the game Côte d'Ivoire (CIV) vs. Zambia (ZAM) played on June 19, 2019, where CIV, the winner and ZAM exchanged 2.21 points. The removal of this game from the \gls{fifa}-recognized list seems to be a reason why \gls{fifa} changed the ratings of both teams between two official publications on Dec. 19, 2019 and on Feb. 20, 2020. Namely, the CIV's rating was changed from 1380 to 1378 and ZAM's from 1277 to 1279. This was done despite both teams not playing at all in this period of time.
\end{itemize}
}

Before discussing the suitable models and algorithms we ask a very simple question: Are the parameters $I_c$ defining the ``importance" of the game suitably set? If not, how should we define them to improve the results? The immediate corollary question is what ``improving'' the results may mean and, in general, how to evaluate the results produced by the algorithm. 

We note here that the concept of the game importance is not unique to the \gls{fifa} rating and appears also in the \gls{fivb} rating, \citep{fivb_rating} and the statistical literature, \eg \citep[Sec.~2.1.2]{Ley19}.

\subsection{Preliminary evaluation of the FIFA rating}\label{Sec:FIFA.vs.FIFA}

A conventional approach in statistics is to base the performance evaluation on a metric, called a scoring function, relating the outcome, $y_t$ to its prediction obtained from the estimates at hand (here, $\btheta_t$), \citep{Gelman14}. 

At this point we want to use only the elements which are clearly defined in the \gls{fifa} ranking and the only explicit predictive element defined in the \gls{fifa} algorithm is the expected score \eqref{FIFA.logistic}, $F(z_t/s)=\Ex[\check{y}_t|z_t]$, we will base the evaluation on the metric affected by the mean. Later we will abandon this simplistic approach.

Using the squared prediction error
\begin{align}\label{squared.error}
    \mfm(z_t, y_t)
    &=
    \big(\check{y}_t-F(z_t/s)\big)^2
\end{align}
averaged over the large number of games,  we obtain the \gls{mse} estimate
\begin{align}\label{MSE.eq}
    \mf{MSE} = \frac{2}{T}\sum_{t=T/2+1}^{T}\mfm(z_t, y_t),
\end{align}
where we use the games in the second half of the observation period to attenuate the initialization effects. This truncation is somewhat arbitrary of course but does not affect the results significantly for large $T$.

The \gls{mse} in \eqref{MSE.eq} may be treated as an estimate of the expectation,
\begin{align}
    \mf{MSE}\approx\Ex_{z_t}\big[\Ex_{y_t|z_t}[\big(\check{y}_t-F(z_t/s)\big)^2]\big]=\Var[\check{y}_t] + \Ex_{z_t}\big[|B(z_t,y_t)|^2\big],
\end{align}
which highlights the bias-variance decomposition, \citep[Ch.~9.3.2]{Duda_book} and 
where  \mbox{$\Var[\check{y}_t]=\Ex_{z_t}\big[\Var[\check{y}_t|z_t]\big]$} is the average conditional variance of $\check{y}_t$, and  $B(z_t,y_t)=F(z_t/s)-\Ex[\check{y}_t|z_t]$ is the estimation bias of the mean.

Therefore, by  reducing the (absolute value of the) bias $B(z_t,y_t)$, that is, by improving the calculation of the expected score $F(z_t/s)$, should manifest itself in a lower value of the \gls{mse}, which is calculated as in \eqref{MSE.eq}.

Using the \gls{mse}, we are now able to assess how the values of the importance parameters $I_c$ (or alternatively, $K$ and $\xi_c$) affect the expected value of the score.  

We find the coefficients $K$ and/or $\xi_c$ by minimizing the \gls{mse} \eqref{MSE.eq} and it turns out that a simple alternate optimization (one variable $K$ or $\xi_c$ is optimized at a time, till convergence) leads efficiently to satisfactory solutions.\footnote{This was done by a line search as we preferred avoiding more formal, \eg gradient-based, methods which are not well suited to deal with the complicated functional relationship resulting from the recursive rating algorithm. } The results are shown in \tabref{tab:solutions.FIFA} and we observe the following:
\begin{itemize}
    \item 
The common update step $K$ increases ten-fold in the optimized setup and it seems that it is the most important contributor to the improvement of the \gls{mse} (which changes from $\mf{MSE}=0.1295$ in the original algorithm to $\mf{MSE}=0.1262$ in the algorithm with fixed-importance games but larger common adaptation step). 
    \item
For the games in the categories well represented in the data, \ie $c\in\set{0,1,2,4,5}$, the relative importance of the games $\xi_c$ does not seem to be critically different and for sure does not fall in line with the values used in the \gls{fifa} algorithm. Overall, the optimized $\xi_c$ yield a very small improvement in the \gls{mse} comparing to the use of fixed $\xi_c$. 

In fact that the Friendlies played in the International Match Calendar window are weighted down ($\xi_1=0.6$) comparing to the Friendlies played outside the window, which is the trend contrary to what the \gls{fifa} algorithm does.  
    \item 
Estimates of $\xi_c$ for the categories $c\in\set{3,6,7,8}$ should not be considered as very reliable because the number of games in each of these categories is rather small (less than $3\%$ of the total). Moreover, the games in the categories $c=7$ and $c=8$ were observed only in June 2018, during the 2018 World Cup; thus, their effect is most likely very weak in the games from the second half of the observed batch, see \eqref{MSE.eq}, which starts around October 2019.
\end{itemize}

\begin{table}[t]
    \centering
    \begin{tabular}{c||c|c|c|c|c|c|c|c|c|c}
    $\mf{MSE}_\tr{opt}$ &
    $K$ & $\xi_0$ & $\xi_1$ & $\xi_2$ & $\xi_3$ & $\xi_4$ & $\xi_5$ & $\xi_6$ & $\xi_7$ & $\xi_8$ \\
    \hline
    $0.1295$ & \ccol    $5$ &\ccol $1$ &\ccol $2$ &\ccol $3$ &\ccol $5$ &\ccol $5$ &\ccol $7$ &\ccol $8$ & \ccol$10$ & \ccol$12$  \\
    $0.1262$ & $12$ &\ccol $1$ &\ccol $2$ &\ccol $3$ &\ccol $5$ &\ccol $5$ &\ccol $7$ &\ccol $8$ & \ccol$10$ & \ccol$12$  \\
    $0.1262$ & $55$ &\ccol $1$ &\ccol $1$ &\ccol $1$ &\ccol $1$ &\ccol $1$ &\ccol $1$ &\ccol $1$ &\ccol $1$ &\ccol $1$  \\
    $0.1250$ & $50$ & \ccol $1$ & $0.6$ & $1.8$ & $0.8$ & $1.2$ & $1.1$ & $2.4$ & $0.1$ & $9.9$  \\
    \end{tabular}
    \caption{Parameters $K$ and $\xi_c$, in \eqref{define.I_c}, are either fixed (shadowed cells), or obtained by minimizing the \gls{mse} \eqref{MSE.eq}. The first row corresponds to the original \gls{fifa} algorithm: $K$ and $\xi_c$ are taken from \tabref{Tab:importance_levels}.}
    \label{tab:solutions.FIFA}
\end{table}

Using a very simple criterion of the \gls{mse} derived from the definitions used by the \gls{fifa} algorithm, we obtain results which cast a doubt on the optimality/utility of the games' importance parameters, $I_c$ proposed by \gls{fifa}.

However, drawing conclusions at this point may be premature. For example, regarding $K$ (which, after optimization should be much larger than $5$), it is possible that the relatively short period of observation time ($29$ months) is not sufficient for small $K$ to guarantee the sufficient convergence but may pay off in a long run, when smaller values of $K$ will improve the performance after the convergence is reached. 
We cannot elucidate this issue with the data at hand.

On the other hand, to address the concerns regarding the \emph{relative} importance weights $\xi_c$ the situation is rather different. Even after the convergence, the weights associated with different game categories should affect meanigfuly the results. To elucidate this point we will now take a more formal approach and go back to the ``drawing board'' to derive the rating algorithm from the first principles. 

\section{Derivation of the algorithm and batch-rating}\label{Sec:derivation.and.batch}

To understand and eventually modify the rating algorithm used by \gls{fifa} we propose to cast it in the well defined probabilistic framework. To this end we define explicitly a model relating the game outcome $y_t$ to the skills of the home-team ($\theta_{i_t}$) and the away-team ($\theta_{j_t}$), where the most common assumption is that the probability that, at time $t$, a random variable $y$  takes the value $y_t$, depends on the skills' difference $z_t=\theta_{t,i_t}-\theta_{t,j_t}$, \ie
\begin{align}\label{Pr.yt}
    \PR{y = y_t|\btheta_t } &= L(z_t/s; y_t)\\
    z_t &= \bx\T_t\btheta_t,
\end{align}
where $L(z_t/s; y_t)$ is the \emph{likelihood} of $\btheta_t$ (for a given outcome $y_t$) and we define a \emph{scheduling} vector $\bx_t=[x_{t,0},\ld, x_{t,N-1}]\T$ for the game $t$, as
\begin{align}
x_{t,m}=\IND{i_t=m} - \IND{j_t=m}, 
\end{align}
with $\IND{a}=1$ when $a$ is true, and $\IND{a}=0$, otherwise. Thus, $x_{t,m}=1$ if the team $m$ is playing at home, $x_{t,m}=-1$ if the team $m$ is visiting, and $x_{t,m}=0$ for all teams $m$ which do not play. This notation allows us to a) deal in a compact manner with all the skills $\btheta_t$ for each $t$, and b) consider the \gls{hfa} or a lack thereof. As before, $s$ is the scale. 

We are interested in the on-line rating algorithms, in which the skills of the participating teams are changed immediately after the results of the game are known. Nevertheless, we will start the analysis with a batch processing, \ie assuming that the skills $\btheta_t$ do not vary in time,  $\btheta_t=\btheta$. This is a reasonable approach if the time window defined $T$ is not too large, so that the skills of the teams may, indeed, be considered approximately constant. The  on-line rating algorithms will be then derived as approximate solutions to the batch optimization problem. The purpose of such approach is to a) tie the algorithm used by \gls{fifa} with the theoretical assumptions, which are not spelled out when the algorithm is presented, b) remove the dependence on the initialization and/or on the scale, and c) treat the past and present data in the same manner, \eg avoiding the partial elimination in the performance metrics, see \eqref{MSE.eq}.

Assuming that the observations are independent when conditioned on the skills, the rating may be based on the \emph{weighted} \gls{ml} estimation principle
\begin{align}
    \hat\btheta 
    \label{ML.optimization}
    &=\argmin_{\btheta}  \sum_{t\in \mcT} \xi_{c_t} \ell(z_t/s;y_t),
\end{align} 
where 
\begin{align}
\label{log.likelihood.def}
    \ell(z_t/s;y_t)
    &=- \log L(z_t/s;y_t),
\end{align} 
is a (negated)\footnote{The negation in \eqref{log.likelihood.def} allows us to use a minimization in \eqref{ML.optimization} which is a very common formulation} log-likelihood. 
The weighting with $\xi_{c_t}\in(0,1]$ is used in the estimation literature to take care of the outcomes which are more or less reliable, \citep{Hu01}, \citep{Amiguet10_thesis}. In our problem, the reliability is associated with the game category, $c_t$, so $\xi_c$ denotes the weight of the category $c$. Since multiplication of $\xi_c$ by a common factor is irrelevant for minimization, we fix $\xi_0=1$.

We may solve \eqref{ML.optimization} using the steepest descent
\begin{align}\label{hat.btheta.gradient}
    \hat\btheta \leftarrow \hat\btheta - \mu/s\sum_{t}\bx_t \xi_{c_t} g(z_t/s;y_t),
\end{align}
where $\mu$ is the adaptation step and 
\begin{align}\label{g.derivative}
    g(z;y) 
    &=  \frac{\dd}{\dd z} \ell(z;y).
\end{align}

The on-line version of \eqref{hat.btheta.gradient} is obtained replacing the batch-optimization with the \gls{sg} which updates the solution each time a new observation becomes available, \ie
\begin{align}\label{SG.algorithm}
    \btheta_{t+1} \leftarrow \btheta_{t} - K\xi_{c_t}\bx_t g(z_t/s;y_t),
\end{align}
where the update amplitude is controlled by the weight $\xi_{x_t}$ and the step $K$ which absorbs the scale $s$.

\subsection{Davidson model and Elo algorithm}\label{Sec:Davidson.Elo}
The rating depends now on the choice of the likelihood function $L(z;y)$ and we opt here for the Davidson model, \citep{Davidson70}, being a particular case of the multinomial model used also in \citet{Egidi21}
\begin{align}
\label{P2.z}
    L(z;\mfH) &= \frac{10^{0.5(z+\eta b)}}{10^{0.5(z+\eta b)}+\kappa+10^{-0.5(z+\eta b)}},\\
\label{P0.z}
    L(z;\mfA) &= P(-z;\mfH),\\
\label{P1.z}
    L(z;\mfD) &= \kappa\sqrt{L(z;\mfH)L(z;\mfA)},
\end{align}
where $\eta$ is a \gls{hfa} modelling the apparent increase in the skills of the local team, the indicator $b=\IND{\tr{game is played in the home-team country}}$ allows us to distinguish between the games played on the home or the neutral venues,\footnote{Out of $T=2964$ games we considered, $768$ were played on neutral venues. To verify the venue we used \citep{roonba} and \citep{soccerway}.} and $\kappa$ controls for the presence of the draws. 

The choice of this model is motivated by the fact that it leads to a simple algorithmic update of the skills generalizing the Elo rating algorithm, \citep{Szczecinski20}. It also becomes equivalent to the latter for a particular value of $\kappa=2$ and $\kappa=0$. These relationships make possible (or at the least -- ease) the comparison with the rating algorithm currently used by \gls{fifa} which is also based on the Elo algorithm.

Using \eqref{P2.z}-\eqref{P1.z} in \eqref{g.derivative}, with straightforward  algebra  we obtain
\begin{align}
\label{g.yz.1}
    g(z;y)
    &=\frac{\dd}{\dd z}\ell(z;y)
    \\
\label{g.yz.3}
    & = -\ln 10(\check{y} - F_{\kappa}(z)),
\end{align}
where $\check{y}$ is the ``score'' of the game which we already defined, and
\begin{align}\label{F.z}
    F_\kappa(z) &= \frac{\frac{1}{2}\kappa+10^{0.5(z+\eta b)}}{10^{0.5(z+\eta b)}+\kappa+10^{-0.5(z+\eta b)}}
\end{align}
has the meaning of the conditional expected score, $F_\kappa(z)=\Ex[\check{y}|z]=\sum_{y\in\mcY} \check{y} L(z; y)$.

Therefore, the \gls{sg} algorithm \eqref{SG.algorithm} becomes 
\begin{align}\label{Elo.algorithm}
        \btheta_{t+1} \leftarrow \btheta_{t} + K\xi_t\bx_t\big(\check{y}_t-F_\kappa(z_t/s)\big)
\end{align}
and it obviously has the form of the Elo and \gls{fifa} rating algorithms, see \eqref{FIFA.basic.rules}-\eqref{FIFA.logistic},  except that we use $F_\kappa(z)$ while the former use $F(z)$. Note that the step $K$ in \eqref{Elo.algorithm} absorbs the term $\ln 10$ from \eqref{g.yz.3}.

It is easy to see that for $\eta=0$ and $\kappa=0$ (\ie when $L(z;\mfD)=0$ and  the draws are ignored) we have  $F_0(z)=F(z)$ which is simply a logistic function as in the Elo algorithm. Furthermore, for $\eta=0$ and $\kappa=2$ we obtain $F_2(z)=F(z/2)$ and thus \eqref{Elo.algorithm} is again equivalent to the Elo rating algorithm but with the doubled scale value.

While we conclude that the \gls{fifa} rating algorithm may be seen as the instance of the maximum weighted likelihood estimation, this is, of course, a ``reverse-engineered'' hypothesis because the \gls{fifa} document, \citep{fifa_rating}, does not mention any remotely similar concept.

\subsection{Regularized batch rating}\label{Sec:regularized.batch}

In order to go beyond the limitation of the \gls{sg} optimization and to avoid the problems related to the removal of the significant portion of the data (meant to eliminate the initialization effects during evaluation, see \eqref{MSE.eq}) we may focus on the original problem defined in \eqref{ML.optimization} for the entire set of data. That is, we ignore now the on-line rating aspect and rather focus on the evaluation of the model and the optimization criterion that underlie the algorithm.

We start noting that the problem \eqref{ML.optimization} is, in general, ill-posed: since the solution depends only on the differences between the skills, $z_t$, all solutions $\hat\btheta$ and $\hat\btheta+\theta_\tr{o}\bone$ are equivalent because the differences $z_t$ are independent from ``origin" value $\theta_\tr{o}$. To remove this ambiguity we may \emph{regularize} the problem as
\begin{align}
    \hat\btheta 
    \label{MAP.optimization}
    &=\argmin_{\btheta} J(\btheta)\\ J(\btheta)&=\sum_{t\in\mcT}\xi_{c_t}\ell(z_t/s;y_t) +\frac{\alpha}{2s^2} \|\btheta\|^2,
\end{align} 
where $\alpha$ is the regularization parameter and we have opted for a so-called ridge regularization \citep[Ch.~3.4.1]{Hastie_book}.

Under the model \eqref{P2.z}-\eqref{P1.z}, the regularized batch-optimization problem  \eqref{MAP.optimization} is  useful to resolve another difficulty. Namely, if there is a team $m$ having registered only wins, \ie when $\forall i_t=m, ~ y_t=\mfH$ and $\forall j_t=m, ~ y_t =\mfA$, then \eqref{ML.optimization} cannot be solved (or rather, $\hat\theta_m\rightarrow\infty$) because $J(\btheta)$ does not limit the value of $\theta_m$. Such a solution not only is unattainable numerically but is, in fact meaningless and the regularization \eqref{MAP.optimization} settles this issue.\footnote{The same problem arises, of course, when a team registers a sequences of pure losses. This is not a hypothetical issue and in the official \gls{fifa} games three teams registered the streaks of unique wins or losses (without any other results): Tonga (three wins), Eritrea (two losses), and American Samoa (four losses). Thus, the attempt to solve the batch-optimization problem without regularization (\ie with $\alpha=0$) would yield $\hat\theta_{m}=\infty$, for $m$ being index of Tonga.}

The estimated skills, $\hat\btheta$ depend now on the weights $\xi_c$, on the regularization parameter $\alpha$, and on the model parameters $\eta$ and $\kappa$. If unknown, all these parameters must be optimized.

As for the optimization criterion, we recall that the \gls{fifa} algorithm only specified the expected score, so the quadratic error \eqref{squared.error} was allowed us to evaluate the algorithm and stay within the boundaries of its definitions. Now, however, with the explicit skills-outcome model, we may go beyond this limitation and may use the prediction metrics known in the machine learning such as the (negated) log-score,  \citep{Gelman14}
\begin{align}\label{log.score.t}
    \mfm^{\tr{ls}}(z_t;y_t) & = \ell(z_t/s;y_t),
\end{align}
often preferred due to its compatibility with the log-likelihood used as the optimization criterion,
or the accuracy score, \citep{Lasek20}
\begin{align}\label{acc.score.t}
    \mfm^{\tr{acc}}(z_t;y_t) & = \IND{y_t = \argmax_{y} L(z_t/s;y)},
\end{align}
which equals one if the event with the largest predicted probability was actually observed, otherwise it is zero.

Furthermore, thanks to the batch-rating we are able to consider the entire data set in the performance evaluation by averaging the scoring function  \eqref{log.score.t} or \eqref{acc.score.t} over all games
\begin{align}
\label{LS.avg}
    \mf{LS}&=\frac{1}{T}\sum_{t\in\mcT} \mfm^{\tr{ls}} \big( \bx_t\T\hat\btheta_{\backslash{t}}, y_t\big),\\
\label{ACC.avg}
    \mf{ACC}&=\frac{1}{T}\sum_{t\in\mcT} \mfm^{\tr{acc}} \big( \bx_t\T\hat\btheta_{\backslash{t}}, y_t\big),
\end{align}
where 
\begin{align}
\label{find.theta}
    \hat\btheta_{\backslash{t}} 
    &= \argmin_{\btheta} J_{\backslash{t}}(\btheta),\\
\label{J.theta.xi.alpha}
    J_{\backslash{t}}(\btheta)
    &=\sum_{\substack{l\in\mcT\\ l\neq t}} \xi_{c_l}\ell(\bx_l\T\btheta/s; y_l) +\frac{\alpha}{2s^2}\|\btheta\|^2.
\end{align}

In plain words, for given parameters ($\alpha$, $\kappa$, $\eta$, $\xi_c$),  we find the skills $\hat\btheta_{\backslash{t}}$ from all, but the $t$-th game [this is \eqref{find.theta}-\eqref{J.theta.xi.alpha}], and next use them to predict the results $y_t$; we repeat it for all $t\in\mcT$, summing the obtained scores. This is the well-known \gls{loo} cross-validation strategy \citep[Sec.~2.9]{Hastie_book}, \citep[Ch.~9.6.2]{Duda_book}: no data is discarded when calculating the metrics \eqref{LS.avg}-\eqref{ACC.avg} and this comes with the price of having to find $\hat\btheta_{\backslash{t}}$ for all $t\in\mcT$. To diminish the computational load, we opt here for the \gls{alo} cross-validation \citep{Rad20} based on the local quadratic approximation of the optimization function defined for all the data. Details are given in \appref{App:ALO}.

Although both, the average log-score in \eqref{LS.avg} and the accuracy \eqref{ACC.avg} can be now optimized with respect to  $\alpha$, $\kappa$, $\eta$, and/or $\xi_c$, we only optimize the log-score whose optimal value is denoted as $\mf{LS}_\tr{opt}$; the resulting accuracy, $\mf{ACC}$ will be also shown. It is, of course, possible to optimize the log-score with respect to any subset of parameters.

Again, we used the alternated minimization: $\mf{LS}$ was minimized with respect to one parameter at a time: $\alpha$, $\kappa$, $\eta$, or $\xi_c$, till no improvement was observed. This simple strategy led to the minimum $\mf{LS}_{\tr{opt}}$ which turned out to be independent of various starting points we used.\footnote{Although we cannot prove the solution to be global, in all our observations the log-score functions seemed to be unimodal.} 

A quick comment may be useful regarding the interpretation of the performance metrics. The accuracy \eqref{ACC.avg} is easily understandable: it is an average number of the events which were predicted correctly (as those with the maximum likelihood $L(z_t/s;y)$). On the other hand, the metric \eqref{LS.avg} may be represented as $\exp(-\mf{LS})=[\prod_{t=1}^T L(z_t/s;y_t)]^{1/T}$ which is a geometric mean of the predicted probabilities assigned to the events which were actually observed. While the accuracy metric penalizes the wrong guesses with zero (so $\mf{ACC}\in[0,1]$), the log-score penalizes them via the logarithmic function, which may be arbitrarily large (so $\mf{LS}\in(0,\infty)$). 

However, the fundamental difference between the two metrics is that we can use the accuracy without specifying the distribution for all possible outcomes but we cannot calculate the log-score in such a case.\footnote{
The common confusion is to interpret the function $F(z_t/s)$ in the Elo/\gls{fifa} algorithm as the probability of the home win, and the value $1-F(z_t/s)$, as the probability of an away win. This, of course, implies that the draw probability equals zero. With such an interpretation, we can still calculate the accuracy metric even if we never predict the draw. On the other hand we cannot calculate the log-score, because when the draw occurs, we have undefined metric $\mf{m}^{\tr{ls}}(z_t/s;\mfD)\rightarrow \infty$.}

\begin{table}[tb]
    \centering
    \begin{tabular}{c||c|c|c||c|c|c|c|c|c|c|c|c||c}
    $\mf{LS}_{\tr{opt}}$ &
    $\alpha$ & $\eta$ & $\kappa$ & $\xi_0$ & $\xi_1$ & $\xi_2$ & $\xi_3$ & $\xi_4$ & $\xi_5$ & $\xi_6$ & $\xi_7$ & $\xi_8$ 
    &  $\mf{ACC} [\%]$\\
    \hline
    $0.960$ &
    $1.7$ &\ccol $0$  &\ccol $2.0$ 
    &\ccol $1$ &\ccol $2.0$ &\ccol $3.0$ &\ccol $5.0$ &\ccol $5.0$ &\ccol $7.0$ &\ccol $8.0$ &\ccol $10.0$ &\ccol $12.0$ 
    &  $55$\\
    $0.948$ &
    $0.2$ &\ccol $0$  &\ccol $2.0$ 
    &\ccol $1$ &\ccol $1.0$ &\ccol $1.0$ &\ccol $1.0$ &\ccol $1.0$ &\ccol $1.0$ &\ccol $1.0$ &\ccol $1.0$ &\ccol $1.0$ 
    &  $56$\\
    $0.948$ &
    $0.3$ &\ccol $0$  &\ccol $2.0$ 
    & \ccol $1$ & $0.9$ & $0.7$ & $0.8$ & $0.9$ & $1.2$ & $0.7$ & $0.8$ & $1.1$ 
    &  $55$\\
    \hline
    $0.918$ &
    $0.3$ & $0.4$  &\ccol $2.0$ 
    &\ccol $1$ &\ccol $1.0$ &\ccol $1.0$ &\ccol $1.0$ &\ccol $1.0$ &\ccol $1.0$ &\ccol $1.0$ &\ccol $1.0$ &\ccol $1.0$ 
    &  $56$\\
    $0.860$ &
    $0.4$ & $0.3$  & $0.8$ 
    &\ccol $1$ &\ccol $1.0$ &\ccol $1.0$ &\ccol $1.0$ &\ccol $1.0$ &\ccol $1.0$ &\ccol $1.0$ &\ccol $1.0$ &\ccol $1.0$ 
    &  $61$\\
    $0.860$ &
    $0.5$ & $0.3$  & $0.8$ 
    & \ccol $1$ & $0.8$ & $0.7$ & $0.8$ & $1.0$ & $1.2$ & $0.8$ & $1.0$ & $1.0$ 
    & $61$\\
    \end{tabular}
    \caption{Batch-rating parameters obtained via minimization of  the log-score \eqref{LS.avg}. The parameters ($\alpha$, $\kappa$, $\eta$, $\xi_c$) are either fixed (shadowed cells), or obtained via optimization. The upper-part results correspond to the conventional \gls{fifa} algorithm: using $\kappa=2$ and $\eta=0$, the expected score is calculated using a logistic function.}
    \label{tab:solutions.batch}
\end{table}

The results obtained are shown in \tabref{tab:solutions.batch} and indicate that
\begin{itemize}
    \item 
The data does not provide evidence for using a category-dependent weights $\xi_c$. There is actually a slight indication that the optimal weights of the Friendlies within the IMC (category $c=1$) and Group phase of Nations Leagues (category $c=2$) are slightly \emph{smaller} than the weight of the regular Friendlies. This stands in contrast to the \gls{fifa} algorithm which doubles the weight $\xi_1$ of the Friendlies played in the IMC and triples the weight of $\xi_2$. 

In fact, the results obtained using the \gls{fifa} weights $\xi_c$ are \emph{worse} than those obtained using constant weights $\xi_c=1$ (\ie essentially ignoring the possibility of weighting). Even with the argument of having a small number of games in some categories (such as a World Cup), it is very unlikely that observing more games will speak in favor of variable weights and almost surely not in favor of the highly disproportionate weights used in the \gls{fifa} algorithm.
    \item
A notable improvement in the prediction capacity as measured by the log-score is obtained by considering the \gls{hfa}. The value $\eta\in \{0.3,0.4\}$ emerges from the optimization fit and we note that $\eta=0.25$ was used in \citet{eloratings.net}.\footnote{Therein, the unnormalized value $\eta s=100$ is reported and since $s=400$, we obtain $\eta=0.25$.}
    \item 
A more important improvement is obtained by optimizing the parameter $\kappa$ which takes into account the draws and their frequency as discussed in \citet{Szczecinski20}. 
\end{itemize}

It is interesting to compare the parameters found by optimization with the simplified formulas proposed in \citet[Sec.~3.2]{Szczecinski20}
\begin{align}
\label{eta.approx}
    \eta &= \log_{10}\frac{f_\mfH}{f_\mfA}\\
\label{kappa.approx}
    \kappa   &= \frac{f_\mfD}{\sqrt{f_\mfH f_\mfA}}  \approx\frac{2 f_\mfD}{1-f_\mfD}.
\end{align}
where $f_y, y \in\mcY$ are empirical frequencies of outcomes. We can consider separately the games played on the neutral venues and calculate these frequencies as $f^\tr{neut.}_{\mfA}=0.37$, $f^\tr{neut.}_\mfD=0.24$, $f^\tr{neut.}_\mfH = 0.39$, and those played on home venues as $f^\tr{hfa}_{\mfA}=0.27$, $f^\tr{hfa}_\mfD=0.22$, $f^\tr{hfa}_\mfH = 0.51$, which yields
\begin{align}
    \kappa^{\tr{hfa}} &= 0.61   &\eta^{\tr{hfa}} &= 0.28\\
    \kappa^{\tr{neut.}} &= 0.63    &\eta^{\tr{neut.}} &= 0.02.
\end{align}

The parameter $\eta^{\tr{hfa}}$ predicted by \eqref{eta.approx} is practically equal to the one obtained by optimization. And while the parameters $\kappa^{\tr{hfa}}$ and $\kappa^{\tr{neut.}}$ are slightly different from the one predicted by \eqref{kappa.approx}, using them in the rating, we obtained $\mf{LS}_{\tr{opt}}=0.868$, which is still notably better than using the conventional \gls{fifa} rating. This is interesting because finding the parameters $\eta$ and $\kappa$ from the frequencies of the games not only avoids optimization but also provides a simple empirical justification.

\section{Margin of victory}\label{Sec:MOV.general}

In the search for a possible improvement of the rating we want to consider now the use of the \gls{mov} variable, defined by the difference of the goals scored by each team, and denoted by $d_{t}$. With that regard, the most recent works adopt two conceptually different approaches. 

The first one keeps the structure of the known rating algorithm (such as \gls{fifa} algorithm) and modifies it by changing the adaptation step size as a function of $d_t$. This was already done in \citet{eloratings.net}, \citet{Hvattum10}, \citet{Silver14}, \citet{Ley19}, and \citet{Kovalchik20}, and is conceptually similar to the weighting according to the game-category we consider in the previous section. 

Second approach changes the model relating the skills to the \gls{mov} variable $d_t$ and was already studied before in \citet{Maher82}, \citet{Ley19}, \citet{Lasek20}, \citet{Szczecinski20c}. We will focus on the simple proposition from \citet{Lasek20} building on the formulation of \citet{Karlis08}.

\subsection{MOV via weighting}\label{Sec:MOV}

For the context, we show in \tabref{tab:goals_difference} the number of games and their percentage of the total, depending on the value of the \gls{mov} variable $d$. While, in principle, it is possible to use directly $d$, it is customary to consider their absolute value, $|d|$. 

The Elo/\gls{fifa} algorithms \eqref{Elo.algorithm} can be easily modified as follows, to take the \gls{mov} variable into account:
\begin{align}
    K_{c,d} = K \xi_{c}\zeta_{d},
\end{align}
where, as before, $K$ is the common step, $\xi_{c}$ is the weight associated with the game-category $c$, and $\zeta_d$ is the function of the \gls{mov}-variable $d$.

\begin{table}[tb]
    \centering
    \begin{tabular}{c|c|c|c|c|c|c}
        $|d|=0$ & $|d|=1$  & $|d|=2$  & $|d|=3$  & $|d|=4$  & $|d|=5$  & $|d|>6$  \\
        \hline
        $678$ & $1070$ & $543$ & $337$ & $162$ & $77$ & $97$\\
         $22\%$ & $36\%$ & $18\%$ & $11\%$ & $6\%$ & $3\%$ & $2\%$
    \end{tabular}
    \caption{\data{Number of games till October 15, 2021} which finished with the goal difference $|d|$ (the fractions are not adding to 100\% due to rounding).}
    \label{tab:goals_difference}
\end{table}

For example, \citep{eloratings.net} uses
\begin{align}
\label{K.d.elorating}
    \zeta_d &=
    \begin{cases}
    1& |d|\le 1\\
    1.5 & |d|=2\\
    1.75 + 0.125(|d|-3) & |d|\ge 3
    \end{cases}.
\end{align}
The similar propositions may be found in \citet{Hvattum10} (in the context of association football), in \citet{Kovalchik20} (to rate the tennis players), or in \citet{Silver14} (for rating of the teams in American football).

To elucidate how useful such heuristics are, we note that the problem is very similar to the importance-weighting we analyzed before; the difference resides in the fact that the weighting depends now on the product $\xi_c\zeta_d$. We may thus reuse our optimization strategy to find the optimal weights for the games with different values of $|d|$. 

To this end we discretize $|d|$ into $V+1$ \gls{mov}-categories, $v=0,\ld, V$ and we use a very simple mapping $v=|d|$ for $v<V$ and $v=V \iff |d|\ge V$. For example, with $V=2$, $\zeta_{0}$ weights the draws ($|d|=0$), $\zeta_1$ weights the games with one goal difference ($|d|=1), v=0,1$ and $\zeta_2$ weights the games with more than one goal difference ($|d|\ge 1$). 

Breaking with the predefined functional relationship as the one shown in \eqref{K.d.elorating} we are more general than the latter, \eg treating the cases $|d|=0$ and $|d|=1$ separately. This makes sense since, not only they are the most frequent events, corresponding, respectively, to $22\%$ and $36\%$ of the total, see \tabref{tab:goals_difference}, but also they correspond to the events of draw and win/loss treated differently by the algorithm.

On the other hand, we are also less general due to the merging of the events $|d_t|\ge V$, although this effect will decrease with $V$, simply because there will be very few observations as may be understood from  \tabref{tab:goals_difference}. For example, with $V=4$, the weighting $\zeta_4$ will be the same for the events with $|d| = 4$ and $|d|>4$ but the latter make only $5\%$ of the total.

We consider again the game categories defined in \tabref{Tab:importance_levels} and thus we solve now the problem
\begin{align}
\label{MAP.optimization.xi.zeta}
    \hat\btheta 
    &=\argmin_{\btheta}  \sum_{t\in\mcT}\xi_{c_t}\zeta_{v_t}\ell(z_t/s;y_t) +\frac{\alpha}{2s^2} \|\btheta\|^2,
\end{align}
where $v_t$ is the index of the \gls{mov} variable $d_t$. To remove ambiguity of the solution, we set $\xi_0=1$ and $\zeta_0=1$.

The parameters $\xi_c$, $\zeta_v$, $\eta$, $\kappa$, and $\alpha$ will be chosen again using the \gls{alo} approach we described in \secref{Sec:regularized.batch}, that is, by optimizing the log-score criterion \eqref{LS.avg}. The results shown in \tabref{tab:solutions.batch.mov} allow us to conclude that:
\begin{itemize}
    \item
    The optimization of the \gls{mov}-weights $\zeta_v$ (while keeping $\xi_c=1$) yields $\mf{LS}_\tr{opt}=0.937$ and the optimization of $\xi_c$ (with $\zeta_v=1$) yields $\mf{LS}_\tr{opt}=0.948$ (see \tabref{tab:solutions.batch}). By comparing them, we see that weighting of the \gls{mov}-categories is more beneficial than weighting of the game-categories. Therefore, there is little improvement in considering the category-related weights, $\xi_c$.
    \item
The optimization indicates that  $\zeta_v$ defined by \eqref{K.d.elorating} is suboptimal. In particular, the optimal \gls{mov} weights, $\zeta_v$ are monotonically growing (as foreseen by the heuristics) only for $|d|\ge 1$ and the draws (\ie $|d_t|=0$) have a weight which is more important that the weights of the events $|d_t|=1$; thus, these two events should not be merged together, nor we should impose a particular functional form for the weights.
    \item 
The best improvement in the prediction is obtained again by optimizing the parameters $\eta$ and $\kappa$ of the Davidson model together with the \gls{mov} weights $\zeta_v$. 
\end{itemize}

\begin{table}[tb]
    \centering
    \begin{tabular}{c|c||c|c|c||c|c|c|c|c|c|c|c||c}
    $\mf{LS}_{\tr{opt}}$ & $V$ &
    $\alpha$ & $\eta$ & $\kappa$ & $\bxi$ & $\zeta_0$ & $\zeta_1$ & $\zeta_2$ & $\zeta_3$ & $\zeta_4$ & $\zeta_5$ & $\zeta_6$
    &  $\mf{ACC} [\%]$\\
    \hline
    $0.949$ & $6$ &
    $0.9$ &\ccol $0$ &\ccol $2.0$ &\ccol $\bone$ 
    &\ccol $1$ &\ccol $1$ &\ccol $1.5$ &\ccol $1.75$ &\ccol $1.875$ &\ccol $2.0$  & $4.2$
    &  $56$\\
    $0.937$ & $6$ &
    $0.2$ &\ccol $0$ &\ccol $2.0$ &\ccol $\bone$ 
    & \ccol $1$ & $0.3$ & $0.5$ & $0.7$ & $1.0$ & $1.3$  & $2.4$
    &  $55$\\
    $0.935$ & $6$ &
    $0.2$ &\ccol  $0$ &\ccol $2.0$ & $\hat{\bxi}$ 
    & \ccol $1$ & $0.3$ & $0.5$ & $0.7$ & $1.0$ & $1.5$  & $2.3$
    &  $55$\\
    \hline
    $0.906$ &   $6$ &
    $0.2$ & $0.4$ &\ccol $2.0$ &\ccol $\bone$
    & \ccol $1$ & $0.3$ & $0.5$ & $0.7$ & $1.0$ & $1.6$  & $3.0$
    &  $56$\\
    $0.852$ &   $6$ &
    $0.3$ & $0.3$ & $0.8$ &\ccol $\bone$ 
    & \ccol $1$ & $0.3$ & $0.5$ & $0.7$ & $1.0$ & $1.6$  & $3.0$
    &  $62$\\
    $0.853$ &   $4$ &
    $0.3$ & $0.3$ & $0.8$ &\ccol $\bone$ 
    & \ccol $1$ & $0.3$ & $0.5$ & $0.8$ & $1.4$ & $\times$  & $\times$
    &  $62$\\
    $0.854$ &   $2$ &
    $0.2$ & $0.3$ & $0.8$ &\ccol $\bone$ 
    & \ccol $1$ & $0.3$ & $0.8$ & $\times$ & $\times$ & $\times$  & $\times$
    &  $62$\\
    $0.857$ &   $1$ &
    $0.2$ & $0.3$ & $0.8$ &\ccol $\bone$ 
    & \ccol $1$ & $0.6$ & $\times$ & $\times$ & $\times$ & $\times$  & $\times$
    &  $61$
    \end{tabular}

    \caption{Batch-rating parameters obtained via minimization of  the log-score \eqref{LS.avg} with weighting of the \gls{mov}-variables. The parameters ($\alpha$, $\kappa$, $\eta$, $\bxi$, $\zeta_v$) are either fixed (shadowed cells), or obtained through optimization; to save space, in the sole case when the parameters $\xi_c$ are optimized, their optimal values are gathered in the vector $\hat\bxi=[1.0, 1.0, 0.9, 1.2, 1.1, 1.8, 1.9, 1.4, 5.6]$.}
    \label{tab:solutions.batch.mov}
\end{table}

\subsection{MOV via modelling}\label{Sec:Skellam.model}

A different approach to deal with the \gls{mov} relies on the integration of the latter into the formal model relating the skills $\btheta_t$ and the observed \gls{mov} variable $d_t$. 

A simple approach proposed in \citet{Karlis08} relies on a direct modelling of the goal difference using the Skellam's distribution 
\begin{align}
\label{Skellam.proba}
    \PR{d_{t}=d|\btheta_t} 
    &=L(z_t;d_t)\\
    &=
    \e^{-(\mu_{\tr{h},t}+\mu_{\tr{a},t})}\left(\frac{\mu_{\tr{h},t}}{\mu_{\tr{a},t}}\right)^{d/2}
    I_{|d|}(2\sqrt{\mu_{\tr{h},t}\mu_{\tr{a},t}}),
\end{align}
where $I_v(t)$ is the modified Bessel function of order $v$ and $\mu_{\tr{h},t}$ and $\mu_{\tr{h},t}$ are means of the Poisson variables modelling the home- and away- goals. The latter are functions of the skills' difference $z_t$, \citep[Sec.~2.2]{Karlis08}
\begin{align}
\label{mu.h.a}
    \mu_{\tr{h},t}& = \e^{c+ z_t + b\eta}, \quad \mu_{\tr{a},t}=\e^{c-z_t-b\eta},
\end{align}
where is $c$ is a constant and, as before, $\eta$ is the \gls{hfa} coefficient.\footnote{For the home team we add-- and for the away team -- subtract $b\eta$ in the exponent. This is different from \citep{Karlis08}, \citep{Lasek20}, where only the home team benefits from the \gls{hfa} boost while the away team is not penalized, see \citep[Eq. (2.2)-(2.3)]{Karlis08}. Of course, we can rewrite \eqref{mu.h.a} as $\mu_{\tr{h},t} = \e^{c'+ z_t + b\eta'}$, $\mu_{\tr{a},t} = \e^{c'+ z_t}$ with $c'=c-b\eta$ and $\eta'=2\eta$ but it makes sense only when the \gls{hfa} is always present, as then $b\eta=\eta$. While this condition holds in the context of football leagues considered in \citet{Karlis08} and in \citet{Lasek20}, this is not the case in the international \gls{fifa} games, which can be played on the neutral venues.}

The model \eqref{Skellam.proba} is a particular case of a more general form shown in \citet{Karlis08}, which allowed us to model the offensive and the defensive skills. Here, however, we are interested in rating and thus one skill per team should be used. As noted in \citet{Ley19}, \citet{Lasek20} this offers a sufficient prediction capability avoiding the problem of over-parametrization due to doubling of the number of skills.

Using \eqref{mu.h.a} in \eqref{Skellam.proba}, the following log-likelihood is obtained
\begin{align}
\label{log.likelihood.Skellam}
    \ell(z;d)
    &=-\log L(z;d)\\
    &=
    (\mu_{\tr{h}}+\mu_{\tr{a}}) - d (z+b\eta)
    - 2\e^c - \log \tilde{I}_{|d|}(2\e^c)
\end{align}
where, for numerical stability it is convenient to use an exponentially modified form of the Bessel function, $\tilde{I}_v(t)=I_v(t)\e^{-t}$, available in many computation packages.

The derivative of \eqref{log.likelihood.Skellam} is given by
\begin{align}
\label{grad.Poisson}
    g(z;d)&=\frac{\dd}{\dd z} \ell(z; d)=-(d - \ov{F}(z)),\\
\label{Ex.score.Poisson}
    \ov{F}(z) &= \mu_{\tr{h}}-\mu_{\tr{a}}=\e^{c}(\e^{z+b\eta}-\e^{-z-b\eta}).
\end{align}

The batch rating consists then in solving the following problem:
\begin{align}
\label{MAP.optimization.Poisson}
    \hat\btheta 
    &=\argmin_{\btheta}  \sum_{t\in\mcT}\ell(z_t/s;d_t) +\frac{\alpha}{2s^2} \|\btheta\|^2
\end{align}
and the \gls{sg} implementation of the \gls{ml} principle will produce the algorithm
\begin{align}\label{Poisson.SG}
    \btheta_{t+1} \leftarrow \btheta_{t} + K\bx_t\big(d_t-\ov{F}(z_t/s)\big),
\end{align}
which is again written in a form similar to the \gls{fifa} rating algorithm, where the goal difference $d_t$ plays the role of the ``score", and $\ov{F}(z_t/s)=\Ex[d_t|z_t]$ is the expected score. The algorithm \eqref{Poisson.SG} can be also obtained by applying the Poisson model to the goals scored by each of the teams \citep{Lasek20}.

To calculate the log-score, we have to merge the events $d<0$ (away-win) and $d>0$ (home-win). Since the closed-form formulas do not exist we do it approximately by truncated sums
\begin{align}
\label{log.score.t.Skellam}
    \mfm^{\tr{ls}}(z;\mfA)&= -\log \sum_{d=-D}^{-1} L(z;d), \quad
    \mfm^{\tr{ls}}(z;\mfD)= -\log L(z;0), \quad
    \mfm^{\tr{ls}}(z;\mfH)= -\log \sum_{d=1}^{D} L(z;d),
\end{align}
where we used $D=50$ which guaranteed that $|1-\sum_{d=-D}^D L(z;d)|<10^{-4}$.

The results shown in \tabref{tab:solutions.batch.Skellam} indicate that, with this very simple approach (with only two parameters of the model which must be optimized) we are able to improve over the \gls{mov}-weighting strategy and this should be attributed to the use of a formal skills-outcome model. The price to pay for the improvement lies in the change of the entire algorithm and in abandoning of the legacy of the Elo algorithm.

Moreover, the possible implementation issues may arise since the expected score \eqref{Ex.score.Poisson} is theoretically unbounded. Whether the improvement of the log-score from $\mf{LS}=0.857$ (in the \gls{mov}-weighting, see \tabref{tab:solutions.batch.mov}) to $\mf{LS}=0.845$ in the Skellam's \gls{mov} model is worth the change and the implementation risks, is at least debatable.

\begin{table}[tb]
    \centering
    \begin{tabular}{c||c|c|c||c}
    $\mf{LS}_{\tr{opt}}$ & $\alpha$ & $\eta$ & $c$ & $\mf{ACC} [\%]$  \\
    \hline
    $0.845$ & $0.21$ & $0.20$ & $0$
    & $61$
    \end{tabular}

    \caption{Batch-rating parameters obtained via minimization of the log-score \eqref{LS.avg} using the Skellam's model \eqref{log.likelihood.Skellam}.}
    \label{tab:solutions.batch.Skellam}
\end{table}

\section{On line rating}\label{Sec:on-line}

\begin{table}[tb]
    \centering
    \begin{tabular}{c|c|c|c}
original    & no shootouts  & no knockouts  & no shootouts\\
            & rules         & rules         &/knockouts rules\\ 
    \hline
BEL (1832.3) & BEL (1831.0) & BRA (1775.9) & FRA (1768.4) \\
BRA (1820.4) & BRA (1817.5) & FRA (1770.1) & BRA (1767.7) \\
FRA (1779.2) & FRA (1778.2) & BEL (1759.2) & BEL (1757.4) \\
ITA (1750.5) & ENG (1740.2) & ITA (1730.5) & ITA (1711.1) \\
ENG (1750.2) & ITA (1733.0) & ENG (1711.9) & ENG (1701.3)
    \end{tabular}
    \caption{Ranking of the top teams: Belgium (BEL), Brazil (BRA), England (ENG), France (FRA), and Italy (ITA). The original \gls{fifa} algorithm and its modified rules are considered.}
    \label{tab:teams.final.ranking}
\end{table}

Before starting a metrics-based comparison of the on-line algorithms, in \secref{Sec:knock.shoot.rules} we will address the use of the knockout/shootout rules \eqref{knockout.rule}-\eqref{shootout.rule} and, in \secref{Sec:Scale.adjustment} the practical issue of setting the scale.

\subsection{Effect of the knockout/shootout rules}\label{Sec:knock.shoot.rules}

\tabref{tab:teams.final.ranking} compares the ranking (of top-five teams) obtained using the \gls{fifa} algorithm (first column) to the rating resulting from the modified algorithm in which we a) eliminate the shootout rule (second column), b) eliminate the knockout rule (third column), as well as c) elimination both rules (fourth column). The differences, most notably the removal of Belgium from the first place, are due to the different number of times the teams benefited from the knockout rules (although the shootout rule for sure has an effect on the final rating too).

Indeed, by analyzing the results of the games, we observed that in the original ranking, Belgium (BEL) benefited four times from the knockout rule for a total of 85 points (which would be lost without the rule \eqref{knockout.rule}), Brazil (BRA) and England (ENG) benefited twice for a total of 53 and 80 points, respectively, while both France (FRA) and Italy (ITA) benefited only once, gaining 14 points each.\footnote{Of course, due to the temporal relationships, eliminating the knockout/shootout rules is not the same as evaluating the points (not lost in the original algorithm) and discarding them from the final results.}

What is important is that the points-preserving knockout rule ignores the direct comparison between the teams. In fact, the games in which Belgium was not penalized (for loosing in knockout stages) were played against France (twice) and against Italy (twice as well). Thus, despite a direct evidence indicating that France and Italy were able to beat Belgium, the knockout rule preserved the points earned by Belgium in other games.

In fact, such a situation is not surprising and we indeed expect the teams which compete for the top ranking spots to be also likely to make it to the final stages of the important competitions (in case of the Belgium's games: World Cup 2018, Euro 2020, and UEFA Nations League 2021) and then play against each other. While these games will provide direct comparison results, current knockout rule will preserve the points of the losing team. 

Whether this is fair and desirable may be debatable especially considering that the knockout/shootout rules are not rooted in any formal modelling principle, and most likely are introduced to compensate for the increased value of $I_c$ in the advanced stages of competitions.

\subsection{Scale adjustment}\label{Sec:Scale.adjustment}

The scale is obviously irrelevant in the batch optimization and the on-line update can also be written in the scale-invariant manner by dividing \eqref{SG.algorithm} by $s$:
\begin{align}
    \btheta'_{t+1} &\leftarrow \btheta'_{t} - K'\xi_{c_t}\bx_tg (z'_t; y_t)\\
    z'_t &= z_t/s\\
    \btheta'_t &= \btheta_t/s\\
    K' &= K/s;
\end{align}
in other words, for the same scale-invariant initialization $\btheta'_0$ and using the same step $K'$ we will obtain the same results $\btheta'_t$.

However, in the \gls{fifa} ranking, a non-zero initialization $\btheta_0$ was determined in advance and thus $\btheta'_0$ is not scale-invariant. 
Thus, given the initialization at hand, the question is how to determine the scale? In general, it is, of course, a difficult question but an insight may be gained assuming that the initialization corresponds to the ``optimal'' solution, \eg $\hat\btheta$ obtained in the batch optimization with a given scale $s_0$. 

It is easy to see that using $s>s_0$ will force the algorithm to significantly change $\btheta_t$ (attainable with large values of the adaptation step, $K$); the same will happen for $s<s_0$ because the optimal estimates $\btheta_t$ will have to be scaled down.

Since scaling up/down of the skills changes their empirical moments we suggest to choose the scale, $s$ in a moment-preserving manner. To this end we define the empirical standard deviation of the skills
\begin{align}
    \sigma_t = \sqrt{\|\hat\btheta_t-\ov{\hat\btheta}_t\|^2/M}    
\end{align}
where $\ov{\hat\btheta}_t=(\sum_{m=1}^M\hat{\theta}_{t,m})/M$ is the empirical mean, and postulate that, at the initialization and at the final step, we have $\sigma_0 \approx \sigma_T$. 

In fact, the initialization used by \gls{fifa} yields $\sigma_0 = 220$ and, after running the \gls{fifa} algorithm we obtain $\sigma_T = 250$, relatively close to the initial value $\sigma_0$.

Changing the scale $s$, we will obtain different $\sigma_T$ so the idea is to run the algorithms for different values of the scale $s$, \eg as multiples of $100$ and to choose the one which yields a standard deviation $\sigma_T \approx \sigma_0$. In practice it has to be done using historical data \emph{before} the new rating is deployed but in our case we could do it in the hindsight.

In this manner we found  $s=200$ to be suitable for the Davidson-Elo algorithm (we obtained \data{$\sigma_T=219$} for the unweighted version and $\sigma_T=221$ for the \gls{mov}-weigted approach), and $s=300$ well suited for the Skellam's algorithms (where \data{$\sigma_T=225$} was obtained). This also indicates the the scale $600$ was too large for the \gls{fifa} rating. This can be noted by comparing, in \tabref{tab:solutions.online.rating} the result \gls{fifa} with $\xi_c=1$ to the results of the \gls{sg} (with $\eta=0$ and $\kappa=2$). Both are essentially the same algorithms (although \gls{fifa} uses the shootout/knockout rules which have negligible impact on the performance) and the only difference resides in the scale. Since the scale $s=200$ in the Elo-Davidson algorithm corresponds to the scale $s=400$ in the \gls{fifa} algorithm, the  latter would perform better with the scale $s=400$. This effect, however, appears only due to limited observation window we have at our disposal and will vanish after a sufficiently large number of games.

\begin{table}[t]
    \centering
    \begin{tabular}{c||c||c|c|c||c}
    algorithm & 
    $\mf{LS}_{\tr{opt}}$ & $K$ & $\eta$  & $\kappa$ & $\mf{ACC} ~[\%]$\\
    \hline
    \gls{fifa}, $\xi_c$ from \tabref{Tab:importance_levels}&
    $0.951$ & 
    \ccol $5$ & \ccol $0$ & \ccol $2$ & 
    $50$\\
    \gls{fifa}, $\xi_c=1$ &
    $0.933$ & 
    $55$ & \ccol $0$ & \ccol $2$ & 
    $52$\\
    \hline
    \multirow{3}{*}{\gls{sg}} &
    $0.917$ & 
    $35$ & \ccol $0$ & \ccol $2$ & 
    $54$\\
     &
    $0.892$ & 
    $35$ & $0.4$ & \ccol $2$ & 
    $58$\\
    &
    $0.841$ & 
    $35$ & $0.3$ &  $0.9$ & 
    $61$
    \end{tabular}
\vskip 0.2cm
    a) Performance of the algorithms : \gls{fifa} ($s=600$) and Elo-Davidson model with \gls{sg} ($s=200$) 
\vskip 0.2cm
    \begin{tabular}{c|c||c|c|c|c|c|c|c||c}
    $\mf{LS}_{\tr{opt}}$ & 
    $V$ & $K$ & $\eta$ & $\kappa$ & 
    $\zeta_0$ & $\zeta_1$ & $\zeta_2$ & $\zeta_3$ & 
    $\mf{ACC}~[\%]$\\
    \hline
    $0.841$ & 
    $1$ & 
    $35$ & $0.3$ & $0.9$ &  
    $1.0$  & $0.9$ & $\times$ & $\times$ &
    $61$\\
    $0.838$ & 
    $2$ & 
    $35$ & $0.3$ & $0.9$ &  
    $1.0$ & $0.6$ & $1.3$ & $\times$ &
    $62$\\
    $0.837$ & 
    $3$ & 
    $40$ & $0.3$ & $0.9$ &  
    $1.0$ & $0.5$ & $0.8$ & $1.8$ & 
    $62$\\
    \end{tabular}
\vskip 0.2cm
    b) \gls{sg} with the \gls{mov} weighting, $s=200$
\vskip 0.2cm
    \begin{tabular}{c||c|c|c||c}
    $\mf{LS}_{\tr{opt}}$ & $K$ & $\eta$ & $c$ & $\mf{ACC} ~[\%]$ \\
    \hline
    $0.827$ & $7.5$ & $0.2$ & $-0.1$ & $62$
    \end{tabular}
\vskip 0.2cm
    c) \gls{sg} implementing Skellam's model for the \gls{mov}, $s=300$
\vskip 0.2cm
    \caption{Parameters and performance of the on-line rating \gls{sg} algorithms obtained by minimizing the log-score \eqref{LS.final} for a) Davidson model, b) \gls{mov}-weighting strategy from \secref{Sec:MOV}, and c) the \gls{mov}-modelling strategy from \secref{Sec:Skellam.model}.}
    \label{tab:solutions.online.rating}
\end{table}

\subsection{Evaluation of the algorithms}\label{Sec:online.algorithms.metrics}

To evaluate the \gls{sg} algorithms for the models studied in the batch context, we will use the log-score and the accuracy metrics defined for the half of the games in the considered time period
\begin{align}\label{LS.final}
    \mf{LS} &= \frac{2}{T}\sum_{t=T/2+1}^{T}\mfm^{\tr{ls}}(z_t, y_t)\\
    \mf{ACC} &= \frac{2}{T}\sum_{t=T/2+1}^{T}\mfm^{\tr{acc}}(z_t, y_t).
\end{align}

We consider the \gls{sg} algorithm based on the Davidson model (\tabref{tab:solutions.online.rating}a), the Davidson model with the \gls{mov}-weighting (\tabref{tab:solutions.online.rating}b), and Skellam's model algorithm (\tabref{tab:solutions.online.rating}c).

In all cases, but in the original \gls{fifa} algorithm, we ignore the category-weighting (\ie we use $\xi_c=1$) because, as we have already shown, its effect is negligible. This is clearly shown in the first part of \tabref{tab:solutions.online.rating}a where we see that using the \gls{fifa} weighting we obtain worse results than when the  weighting in ignored. This is essentially the same result as the one we have shown in \tabref{tab:solutions.FIFA} but we repeat it here to show the log-score metric which we could not calculate without first introducing the Davidson model underlying the \gls{fifa} algorithm.

The results indicate that:
\begin{itemize}
    \item 
The most notable improvements are due to, in similar measures, two elements: the introduction of the \gls{hfa} coefficient, $\eta$ and the explicit use of the Davidson model (and thus, the optimization of the coefficient $\kappa$).
    \item
Additional small, but still perceivable gains are obtained by introducing the \gls{mov}-weighting, where from the lesson learnt in \secref{Sec:MOV} we weight independently the draws and the home/away wins.
    \item
The \gls{mov}-modelling using the Skellam's distribution brings again a small benefit.
\end{itemize}

We present in \tabref{tab:teams.final.ranking.new} the rating obtained for the top teams via new rating algorithms. Of course, due to smaller scale we used, the skills have smaller values and should not be compared directly to those from \tabref{tab:teams.final.ranking} but the ranking is of interest, where the teams from the \gls{fifa} ranking are present (FRA, BRA, BEL) but this time Argentine (ARG), which was on the sixth place in the previous rankings, is now consistently on and above the top-third position. We can also see that the differences between the rating values are much less pronounced.

\begin{table}[tb]
    \centering
    \begin{tabular}{c|c|c}
Davidson    & MOV weights  & Skellam's model\\
\hline
FRA (1683.5) & FRA (1690.8) & BRA (1596.0) \\
BRA (1673.1) & BRA (1677.4) & ARG (1585.8) \\
ARG (1668.6) & ARG (1677.0) & BEL (1546.2) \\
BEL (1664.9) & BEL (1666.5) & POR (1541.2) \\
ITA (1657.7) & ITA (1665.6) & ESP (1540.7) 
    \end{tabular}
    \caption{Ranking of the top teams using the proposed algorithms.}
    \label{tab:teams.final.ranking.new}
\end{table}

\section{Conclusions}\label{Sec:Conclusions}

In this work we analyzed the \gls{fifa} ranking using the methodology conventionally used in the probabilistic modelling and inference. In the first step, we identified the model relating the outcomes (games results) to the parameters which have to be optimized (skills of the teams). More precisely, we have shown that the \gls{fifa} algorithm can be formally derived as the \acrfull{sg} optimization of the  weighted \acrfull{ml} criterion in the Davidson model \citep{Davidson70}.

This first step allows us to define the performance metrics related to the predictive performance of the algorithms we study. This is particularly important in the case of the \gls{fifa} ranking algorithm because it does not model the outcomes of the game but only explicitly specifies the expected score, which is not sufficient to precisely evaluate the rating results. It also allows us to apply the batch approach to rating and skills' estimation. This conventional machine learning strategy frees us from the considerations related to the scale, initialization, or modeling of the skills' dynamics.

Using the batch rating, we have shown that the game-category weighting is negligible at best, and counterproductive at worst, which is the case of the weighting used by the \gls{fifa} rating. This observation is interesting in its own right because, while on one hand the concept of weighting is also used in the rating literature, \eg \citep{Ley19}, on the other, the literature does not show any evidence that it is in any way beneficial and our findings consistently indicate the contrary.

We next considered extensions of the algorithm by including the \gls{hfa} and optimizing the parameter responsible for the draws. These two elements seem to be particularly important from the point of view of the performance of the rating algorithm. While the \gls{hfa} is a well-known element, already considered by \gls{fifa} in \citet{fifa_rating_W}, the possibility of generalizing the Elo algorithm by using the Davidson's model, was only recently shown in  \citet{Szczecinski20}. 

We also evaluated the possibility of using the \acrfull{mov} given by the goal differential, where we analyzed the weighting strategy and the modelling based on the Skellam's distribution. These two methods further improve the results at the cost of higher complexity. Here, the formal optimization strategy of the weighting parameters also yield interesting and somewhat counter-intuitive results. Namely, we have shown that the games won with small margin should have smaller weights than the tied games. This stands in net contrast with the weighting strategies proposed before, \eg in \citet{Hvattum10}, \citet{Silver14}, \citet{Kovalchik20} which use the weighting with monotonically increasing functions of the margin.

Finally, we evaluated the heuristic shootout/knockout rules which are used in the \gls{fifa} rating. Since their impact on the overall performance is small and they may distort the relationship between the ratings of the strong teams which often face each other in the final stages of the competitions, their usefulness is questionable. In particular, eliminating the knockout rule would strip Belgium from its first place position in the current \gls{fifa} ranking due to  multiple losses Belgium suffered against the current top teams (\eg Italy, France).

\subsection{Recommendations}\label{Sec:Recommedations}

Given the analysis and the observations we made, if the \gls{fifa} rating was to be changed, the following steps are recommended:
\begin{enumerate}
    \item 
Add the \acrfull{hfa} parameter to the model because playing at the home venue is a strong predictor for the victory. Not only this well-known fact is already exploited in Women \gls{fifa} ranking but such a modification is most likely the simplest and the least debatable element. In our view, it is surprising that the current rating adopted in 2018 does not include the \gls{hfa}.
    \item
Use explicit model to relate the skills to the outcomes. Not only it would add expressiveness providing the explicit predicted probability for each outcomes, but it also improves the prediction results. Note that the rating algorithm introduced recently by \gls{fivb} adopts such an approach and specifies the probability for each of the game outcomes. In the context of the \gls{fifa} ranking, the Davidson model we used in this work is an excellent candidate for that purpose as it results in a natural generalization of the Elo algorithm, preserving the legacy of the current algorithm. 
    \item 
Remove the weighting of the games according to their assumed importance because the data does not provide any evidence for their utility, or rather provides the indication that the weighting in its current form is counterproductive. If the concept of the game importance is of extra-statistical nature (such as entertainment), it is preferable to diminish its role, \eg by shrinking the gap between the largest and the smaller values of $\xi_c$ used.
    \item 
Remove the shootout and knockout rules which are not rooted in any sound statistical principle.

As far as the knockout rule is concerned, while the intent to protect the rating of the teams which manage to qualify to the knockout stage is clear, we may argue that the penalty due to losing in the knockout game is aggravated by the increased weighting of these games. Therefore, removing the weight, as we postulate, would also eliminate the very reason to protect the teams' points with the knockout rule.

Regarding the shootout rule, a small frequency of events where it can be applied and a marginal changes in the score imposed by the rule, make its impact rather negligible. Its fairness is again debatable because there is little evidence relating the skills of the teams to the outcome of the shootout.
    \item 
If the rating was to consider the \gls{mov}, the simplest solution lies in weighting the update step using the goal differential. On the other hand, the modification based on the change of the model using the Skellam's distribution may cause numerical problems and the relatively small performance gains hardly justify the added complexity.

On the other hand, the \gls{mov} may be added using alternative solutions similar to those already considered in the Women' teams \gls{fifa} ranking. Again, the latter should be studied, \eg using the methodology we used in this work and basing the results on a formal probabilistic model.
\end{enumerate}

\appendix
\numberwithin{equation}{section}
\renewcommand{\theequation}{\thesection.\arabic{equation}}
\section{Approximate leave-one-out cross-validation}\label{App:ALO}

Our goal is to calculate in a simple manner the terms $\bx_t\T\hat\btheta_{\backslash{t}}$ which appear in the scoring function in \eqref{LS.avg} and in \eqref{ACC.avg}.

We start by approximating the \gls{map} objective function \eqref{J.theta.xi.alpha} using the Taylor series
\begin{align}
    J_{\backslash{t}}(\btheta)
    &=J(\btheta)+\xi_{c_t}
    \log L(\bx_t\T\btheta/s; y_t)\\
\nonumber
    &\approx 
        J(\hat\btheta) + \xi_{c_t}
    \log L(\bx_t\T\hat\btheta/s; y_t) \\
\label{J.slash.t.approx}
    &\quad
    -\frac{\xi_{c_t}}{s}g_t\bx_t\T(\btheta-\hat\btheta)
    + \frac{1}{2}(\btheta-\hat\btheta)\T 
    \Big[
    \hat\matH - \frac{\xi_{c_t}}{s^2}h_t\bx_t\bx_t\T
    \Big] 
    (\btheta-\hat\btheta)
\end{align}
where $g_t \equiv g(\bx\T_t\hat\btheta/s;y_t)$ is defined in \eqref{g.yz.3},
\begin{align}
\label{find.theta.0}
    \hat\btheta &= \argmin_{\btheta} J(\btheta)
\end{align}
is the optimal solution for all data, the Hessian at optimum is given by
\begin{align}
    \hat\matH 
    &= \nabla^2_{\btheta}J(\btheta)|_{\btheta=\hat\btheta}
    =
    \sum_{t\in\mcT} \frac{\xi_{c_t}}{s^2}h_t\bx_t\bx\T_t + \frac{\alpha}{s^2}\bI,
\end{align}
and we use second derivative $h_t\equiv h(\bx\T_t\hat\btheta/s)$ with  \citep[Sec.~IV]{Szczecinski21}
\begin{align}
\label{Davidson.h}
    h(z)&=\frac{\dd}{\dd z} g(z;y)
    = 
     \frac{\left(\ln 10\right)^2}{4}\frac{\kappa 10^{0.5(z+\eta b)}+4 +\kappa 10^{-0.5(z+\eta b)}}{(10^{0.5(z+\eta b)}+\kappa+ 10^{-0.5(z+\eta b)})^2}.
\end{align}

By equating the gradient of  \eqref{J.slash.t.approx} to zero, we find the approximate solution to the optimization problem
\begin{align}
\nonumber
    \hat{\btheta}_{\backslash{t}}
    &\approx 
    \argmin_{\btheta} {J}_{\backslash{t}}(\btheta)\\
\label{theta.hat.slash.t}
    &=
    \hat\btheta +\frac{\xi_{c_t}g_t}{s} \Big[
    \hat\matH - \frac{\xi_{c_t}}{s^2}h_t\bx_t\bx_t\T
    \Big]^{-1}\bx_t
\end{align}
and the terms $\bx_t\T\hat\btheta_{\backslash{t}}, t\in\mcT$ which appear as arguments of the metrics \eqref{LS.avg} and \eqref{ACC.avg} can be now calculated efficiently for all $t\in\mcT$ once $\hat\btheta$ is known \citep{Rad20}\citep{Burn20}
\begin{align}\label{x.theta.t}
    \bx_t\T\hat{\btheta}_{\backslash{t}}
    &\approx
    \bx_t\T\hat\btheta 
    +\frac{\xi_{c_t}g_t}{s} \bx_t\T\Big[
    \hat\matH - \frac{\xi_{c_t}}{s^2}h_t\bx_t\bx_t\T
    \Big]^{-1}\bx_t
    \\
    \label{x.theta.t.2}
    &=
    \bx_t\T\hat\btheta 
    +\frac{\xi_{c_t}g_t}{s} \bx_t\T\Big[
    \hat\matH^{-1} + \frac{\xi_{c_t}h_t}{s^2-\xi_{c_t}h_t\bx_t\T\hat\matH^{-1}\bx_t}
    \hat\matH^{-1}\bx_t\bx_t\T\hat\matH^{-1}
    \Big]\bx_t
    \\
    \label{x.theta.t.3}
    &=
    \bx_t\T\hat\btheta  + \frac{\xi_{c_t}g_t a_t s}{s^2 - \xi_{c_t} h_t a_t},
\end{align}
where $a_t=\bx_t\T\hat\matH^{-1}\bx_t$ and to pass from \eqref{x.theta.t} to \eqref{x.theta.t.2} we used the matrix inversion lemma  \citep[Ch.~A.1.8]{Barber12_Book}.

The advantage of this formulation is clear: instead of solving $T$ times the optimization problem \eqref{find.theta}, we only need to solve once the optimization defined in \eqref{find.theta.0}. In comparison with the latter, the remaining operations of the inversion of the matrix $\matH_0$ and the multiplication required to calculate $a_t, t\in\mcT$, have a very small complexity.

The identical approach may be used to apply the \gls{alo} to the problem \eqref{MAP.optimization.xi.zeta} but, we have to replace  $\xi_{c_t}$ in  \eqref{x.theta.t.3} with $\xi_{c_t}\zeta_{v_t}$.

In order to apply the \gls{alo} to the problem  \eqref{MAP.optimization.Poisson} we need a second derivative of \eqref{grad.Poisson} which is given by
\begin{align}
    h(z) =\frac{\dd}{\dd z} g(z;d) 
    &= \e^{c}(\e^{z+b\eta}+\e^{-z-b\eta}).
\end{align}


\end{document}